\documentclass[sn-mathphys,Numbered]{sn-jnl}% Math and Physical Sciences Reference Style
\usepackage{graphicx}%
\usepackage{multirow}%
\usepackage{amsmath,amssymb,amsfonts}%
\usepackage{amsthm}%
\usepackage{physics}%
\usepackage{mathrsfs}%
\usepackage[title]{appendix}%
\usepackage{xcolor}%
\usepackage{textcomp}%
\usepackage{manyfoot}%
\usepackage{booktabs}%
\usepackage{algorithm}%
\usepackage{algorithmicx}%
\usepackage{algpseudocode}%
\usepackage{listings}%
\usepackage{siunitx}

\theoremstyle{thmstyleone}%
\theoremstyle{thmstyletwo}%

\theoremstyle{thmstylethree}%

\begin{document}

%\title[Article Title]{Nonreciprocal spin dynamics in 3D artificial chiral magnets with magnetic helices at room temperature}
%\title[Article Title]{Room temperature realization of artificial chiral magnets with reprogrammable magnon nonreciprocity at zero field}
\title[Article Title]{Room temperature realization of artificial chiral magnets with reprogrammable magnon nonreciprocity at zero field}

%%=============================================================%%
%% Prefix	-> \pfx{Dr}
%% GivenName	-> \fnm{Joergen W.}
%% Particle	-> \spfx{van der} -> surname prefix
%% FamilyName	-> \sur{Ploeg}
%% Suffix	-> \sfx{IV}
%% NatureName	-> \tanm{Poet Laureate} -> Title after name
%% Degrees	-> \dgr{MSc, PhD}
%% \author*[1,2]{\pfx{Dr} \fnm{Joergen W.} \spfx{van der} \sur{Ploeg} \sfx{IV} \tanm{Poet Laureate} 
%%                 \dgr{MSc, PhD}}\email{iauthor@gmail.com}
%%=============================================================%%

\author*[1]{\fnm{Mingran} \sur{Xu}}\email{mingran.xu@epfl.ch}
\equalcont{These authors contributed equally to this work.}
\author[1]{\fnm{Axel} \sur{J. M. Deenen}}
%\email{axel.deenen@epfl.ch}
\equalcont{These authors contributed equally to this work.}
\author[1]{\fnm{Huixin} \sur{Guo}}
%\email{huixin.guo@epfl.ch}
% \equalcont{These authors contributed equally to this work.}
\author*[1,2]{\fnm{Dirk} \sur{Grundler}}\email{dirk.grundler@epfl.ch}
% \equalcont{These authors contributed equally to this work.}

\affil*[1]{\orgdiv{Laboratory of Nanoscale Magnetic Materials and Magnonics, Institute of Materials (IMX)}, \orgname{School of Engineering, École Polytechnique Fédérale de Lausanne (EPFL)}, \orgaddress{\city{Lausanne}, \postcode{1015}, \state{Vaud}, \country{Switzerland}}}

\affil[2]{\orgdiv{Institute of Electrical and Micro Engineering}, \orgname{School of Engineering, École Polytechnique Fédérale de Lausanne (EPFL)}, \orgaddress{\city{Lausanne}, \postcode{1015}, \state{Vaud}, \country{Switzerland}}}

%%==================================%%
%% sample for unstructured abstract %%
%%==================================%%

%word limit 150

\abstract{
%\textbf{word limit 150}
Chiral magnets are materials which possess unique helical arrangements of magnetic moments, which give rise to nonreciprocal transport and fascinating physics phenomena. On the one hand, their exploration is guided by the prospects of unconventional signal processing, computation schemes and magnetic memory. On the other hand, progress in applications is hindered by the challenging materials synthesis, limited scalability and typically low critical temperature. Here, we report the creation and exploration of artificial chiral magnets (ACMs) at room temperature. By employing a mass production compatible deposition technology, we synthesize ACMs, which consist of helical Ni surfaces on central cylinders. Using optical microscopy, we reveal nonreciprocal magnon transport at GHz frequencies. It is controlled by programmable toroidal moments which result from the ACM’s geometrical handedness and field-dependent spin chirality. We present materials-by-design rules which optimize the helically curved ferromagnets for 3D nonreciprocal transport at room temperature and zero magnetic field.}


%%\pacs[JEL Classification]{D8, H51}

%%\pacs[MSC Classification]{35A01, 65L10, 65L12, 65L20, 65L70}

\maketitle

\section{Introduction}\label{sec1}

%% chiral magnets must have PT breaking but PT breaking not necessary give Chiral magnets.
%% magnets with torodial moment not necessay to be Chiral, for instance, Neel domain wall, whichi is nonchiral but with non-zero toroidal moment
%% Chiral magnet must has toroidal moment, but not necessay parallel to the chiral axis, for instance, Bloch domain is chiral with toroidal moment but Helical spin is chiral but with compensated toroidal moment (zero), but it still can induce linear magnetoelectric effect.
%% 

Chirality is a characteristic quality of objects which are non-superimposable mirror images of each other. It plays a crucial role in nature, various scientific disciplines and technologies~\cite{Barron1972,Rikken1997,He2023,Eslami2014}. Manifested by nonreciprocal transmission of signals~\cite{Rikken2002,Cheong2018}, chiral objects manipulate the polarization of electromagnetic waves. In communication technologies, chirality is being widely integrated for signal isolation, noise reduction and efficient energy transfer. In condensed matter physics and materials science, chiral crystals form a fascinating class of magnetic materials, which give rise to rich physical phenomena. Among them, there are the nonreciprocal propagation of phonons ~\cite{Nomura2019} and magnons~\cite{Seki2016a}, skyrmions as topologically protected spin structures, the magnon Hall effect~\cite{Onose2010}, bulk electronic diode effect~\cite{Yokouchi2017} and emergent inductor~\cite{Yokouchi2020}. Magnetic chirality arises from breaking spatial-inversion and time-reversal symmetries simultaneously~\cite{Barron2000,Cheong2022a}, which occurs  in specific noncentrosymmetric crystals~\cite{Tokura2018}. Typically, the crystals require complex synthesis routes and possess the relevant magnetic ordering at low temperatures. These aspects delay the development of room-temperature applications.

\begin{figure}[h]%
\centering
\includegraphics[width=1\textwidth]{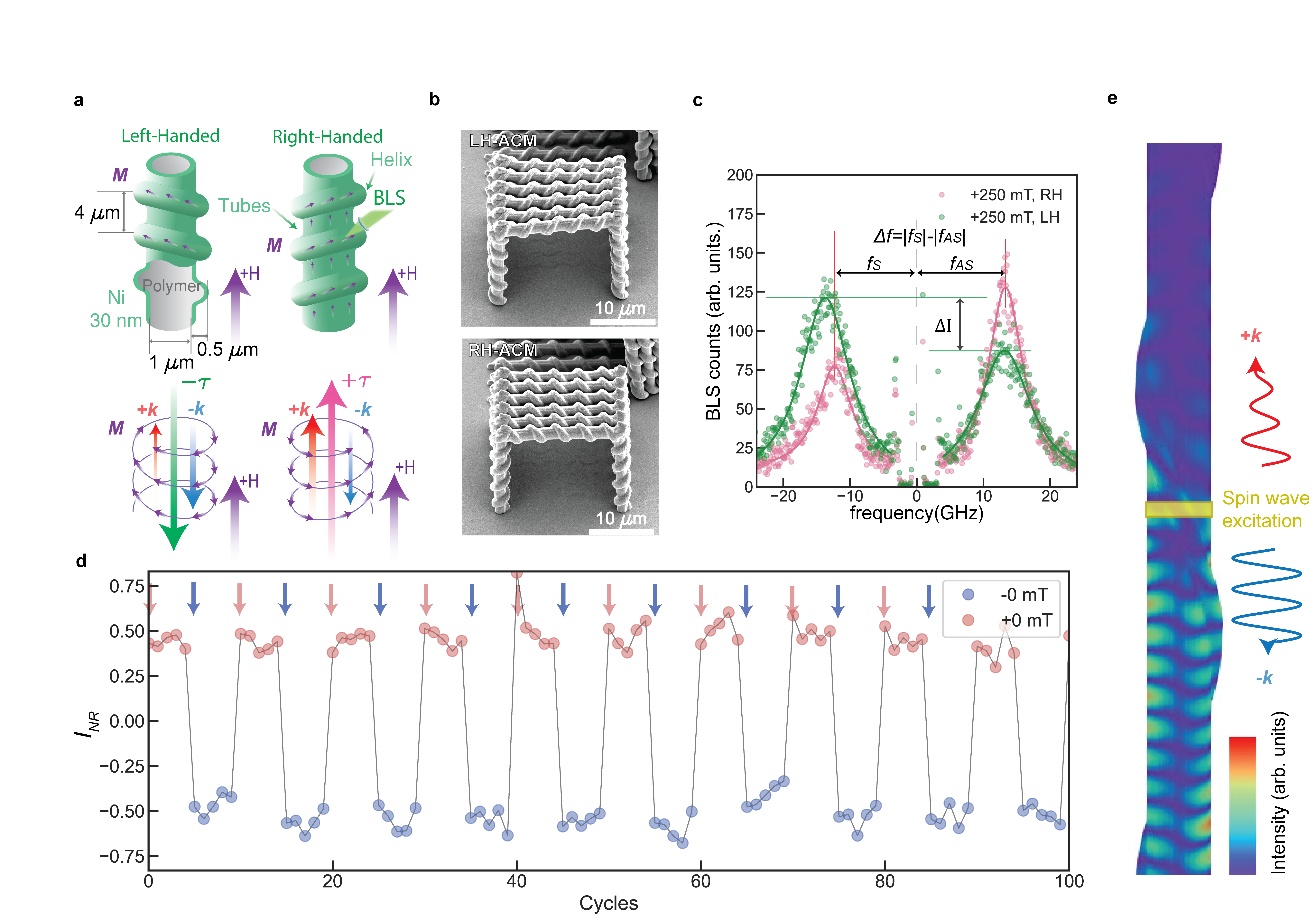}
\caption{\textbf{Helical spin texture of artificial chiral magnets, toroidal moments and reprogrammable nonreciprocity of magnons at zero magnetic field.} \textbf{a,} Sketch of ferromagnetic tubes with helical magnetic states of left- and right-handedness in a small magnetic field $H$ (top) and corresponding toroidal moment vectors $\boldsymbol{\tau}$ motivating nonreciprocal (NR) propagation of magnons with wave vectors $+k$ and $-k$ along $z$-direction (bottom). \textbf{b,} Ferromagnetic tubes incorporating screw-like surfaces with left-handedness and right-handedness prepared by additive manufacturing. The surface topology of the Ni shell gives rise to an artificial chiral magnet (ACM). \textbf{c,} Magnon spectra obtained on a RH-ACM (pink symbols) and LH-ACM (green symbols) in a field $\mu_0H=+250$~mT at position $z=-0.5$ \SI{}{\um}. Curves reflect Lorentzian fits. \textbf{d,} Nonreciprocity parameter $I_{\rm NR}$ extracted from spectra taken at $\mu_0H=0$~mT after applying a magnetic field of 30~mT in opposite axial directions of a LH-ACM. $I_{\rm NR}$ was measured at $f=10~$GHz five times (cycles) before realizing a reversed magnetic field history at position $z= -0.7$ \SI{}{\um}. The data reflect reprogrammable nonreciprocity at zero field. \textbf{e,} Micromagnetic simulation revealing nonreciprocal magnon transport into negative $z$-direction when exciting an LH-ACM in its central region at zero field (here, at 3 GHz) after it has been exposed to a field pointing in positive $z$-direction beforehand.}\label{fig1}
\end{figure}

In recent years, the advancements in nanotechnologies~\cite{Saha2019} have opened the possibility of endowing conventional materials with complex topologies in all three spatial directions ~\cite{Donnelly2017}. They allow one to artificially design optical~\cite{Wang2023}, mechanical, and electrical properties. In magnetism, the interplay between 3D geometry and magnetic properties is intensively investigated~\cite{Fernandez-Pacheco2017,Makarov2022,Guo2023a,Volkov2024a,Goebiewski2024}. It has been found that curved surfaces and geometric topology lead to the emergence of an extrinsic Dzyaloshinskii-Moriya interaction (DMI)~\cite{Volkov2018,Makarov2022b}, chiral symmetry breaking and unconventional magnetic anisotropies. For spin waves (magnons), nonreciprocal band structures are expected. Nonreciprocity means that the energy flow is different in opposite directions. In a nonreciprocal magnonic device, either the signal intensity, the frequency, or both differ for magnons propagating with opposite wave vectors $-k$ and $+k$. For bulk crystals and thin films nonreciprocity was examined by means of broadband electrical spectroscopy with coplanar waveguides~\cite{Seki2016a}, magnetoelastic spectroscopy with surface acoustic waves~\cite{Sasaki2017,Xu2020}, and Brillouin light scattering (BLS) spectroscopy with wave vector resolution~\cite{Di2015, Nembach2015,Ogawa2021}, confocal microscopy~\cite{Madami2011} or time-resolved scanning transmission X-ray microscopy (TR-STXM)~\cite{Gallardo2023}. The symmetry breaking and magnon nonreciprocity arose from interfacial DMI provided through an additional heavy metal layer, magnetic surface anisotropy~\cite{Hillebrands1990,Cochran1990}, dipolar coupling and asymmetric coupling of hybridized modes~\cite{Yamamoto2020}.  In a more general picture, the spatial-symmetry breaking of a magnetic system via e.g. a heavy-metal interface, antiferromagnetically coupled bilayers, a Bloch domain wall, or curved shapes generates a non-zero toroidal moment vector $\boldsymbol{\tau}=\sum_i \boldsymbol{r}_i \times \boldsymbol{m}_i$ with $\boldsymbol{r}_i$ ($\boldsymbol{m}_i$) the position vector (magnetic moment) at site $i$ (Fig. \ref{fig1}a).
It is used to estimate the helicity $\chi$ of a magnetic configuration via integration across its cross-sectional area $A$ using~\cite{Korber2022a}
\begin{equation}
\centering
\boldsymbol{\tau}\left(\boldsymbol{m}_0\right) \stackrel{\text { def }}{=} \frac{1}{A} \iint_A \mathrm{~d} x \mathrm{d} y \boldsymbol{~r} \times \boldsymbol{m}_0(\boldsymbol{r}),
\label{eq:tormom}
\end{equation}
where $\boldsymbol{m_{0}}$ represents the static magnetization. If the relation
\begin{equation}
\centering
\boldsymbol{k} \cdot \boldsymbol{\tau(\boldsymbol{m}_0)} \neq 0
\label{eq:nonrec_toroidal}
\end{equation}
is fulfilled, the spin-wave dispersion relation exhibits nonreciprocity which is induced by dipolar interaction between the spins ~\cite{Korber2022a}. 
Via non-zero $\tau$ and the dipole-induced symmetry breaking a helical spin structure, one can introduce chirality in a conventional but curved magnet. Since the beginning of this decade, considerable efforts have been dedicated to investigating the spin dynamics of curved ferromagnetic shells forming nanotubes \cite{Giordano2020,Korber2021a,Giordano2023a} and lattices of interconnected ferromagnetic curved segments ~\cite{Guo2023a, Goebiewski2024}. However, a conclusive experimental evidence of NR magnon propagation has remained elusive as noted in Ref.  \cite{Korber2021a}. Helicity and chirality are two closely related physical concepts but refer to different characteristics of an object. A helical spin texture stands for spins rotating relative to an axis in space, while a chiral spin texture describes a non-superimposable arrangement of spins. In a bulk chiral magnet, spins relax into the helical texture at zero magnetic field. Its toroidal moment vanishes~\cite{Spaldin08}, and the chiral magnet does not exhibit an asymmetric dispersion relation in zero field~\cite{Ogawa2021}.

Beyond the intriguing physics emerging from curved surfaces, 3D nanomagnetic structures captivate vast attention from a technological perspective. Compared with planar thin-film devices, 3D magnetic architectures innately come with a greater data storage capacity and higher dimensional connectivity. These aspects are key for sustainable nanoelectronics with low energy consumption and high computational power. Towards magnonics-based signal processing, nonreciprocal waveguides are of utmost importance for an efficient routing of signals. Furthermore, artificially created 3D magnetic objects naturally have a higher degree of design flexibility. If prepared from magnetically ordered materials with high critical temperature the materials-by-design approach is expected to enable chiral properties ready for room-temperature applications.  

Here, we present a novel route to chirality in magnetic materials at room temperature. Using the materials-by-design approach and additive manufacturing of nickel, we create artificial chiral magnets (ACMs) and explore their unprecedented magnetic properties by means of inelastic light scattering and simulations. By combining two-photon lithography (TPL) and mass production compatible atomic layer deposition (ALD), we prepare tubular ferromagnetic structures with screw-like surface topology of left-handedness (LH) and right-handedness (RH). The ACMs consist of a photo-resist template (illustrated in gray color in Fig.  \ref{fig1}a,) which is covered by a uniformly thick nickel shell (green). A scanning electron microscopy image of LH-ACMs is shown in Fig.~\ref{fig1}b. The suspended magnetic tubes (bright) are investigated in the present work.
The helical relief (helix region) is integrated into an otherwise cylindrical Ni tube (tube region). In a magnetic field $H$ applied parallel to the ACM, the magnetization in the tube region aligns with $H$ while in the helix region it spirals around the tube forming locally a screw-like spin texture. The handedness of the texture follows the corresponding structural chirality, generating oppositely polarized toroidal moments $\mp \boldsymbol{\tau}$ in LH- and RH-ACMs and inverted nonreciprocities $\boldsymbol{\tau} \cdot \boldsymbol{k}$ (Fig.  \ref{fig1}a). By means of micro-focused Brillouin light scattering microscopy ($\mu$-BLS) (Fig.  \ref{fig1}a and c), we detect nonreciprocal spin dynamics with high spatial resolution. By studying Stokes and anti-Stokes scattering processes of the focused laser light with magnons, we identify differences in both resonance frequencies $\Delta f$ (Fig.  \ref{fig1}c) and intensities of BLS peaks (Fig.  \ref{fig1}d) which we describe by the asymmetry parameter $I_{\rm NR}$. We anticipate that the observed nonreciprocities originate from asymmetric dispersion relations induced by chiral spin textures following the helical reliefs. By scanning the laser spot across the curved surfaces of the helices, we resolve a gradual evolution of Stokes and anti-Stokes peaks. We attribute the variations to momentum conservation dependent on the light scattering geometry. The nonreciprocities are inverted by either changing the polarity of the applied magnetic field or exploring an ACM of opposite handedness. We demonstrate also a programmable magnetic chirality at zero field through the magnetic field history applied to the helical Ni tubes. Exploring the remanent state of the same tube after applying either +30 or $-$30 mT, we detect opposite nonreciprocities at zero magnetic field (labelled $\pm 0$~mT) by means of $\mu$-BLS (Fig. \ref{fig1}d). Micromagnetic simulations show that the surface curvature of the helical relief added to the Ni tubes stabilizes a chirally arranged magnetic texture. The observed signatures of nonreciprocity are qualitatively reproduced by numerical simulations (Fig. \ref{fig1}e). Hence, our research unveils the possibility of empowering a conventional single-layered ferromagnet such as Ni with chiral properties in the remanent state through engineered helical surface curvature. Using an additive manufacturing strategy to create the observed nonreciprocity of spin dynamics, we facilitate a mass production technology for chirality-based nanoelements which offer unidirectional signal transmission in 3D nanomagnetic architectures. We expect them to enhance connectivity in hardware-implemented 3D neural networks. 

% time c

\section{Observation of asymmetric magnon spectra in artificial chiral magnets}\label{sec2}
\begin{figure}[h]%
\centering
\includegraphics[width=1\textwidth]{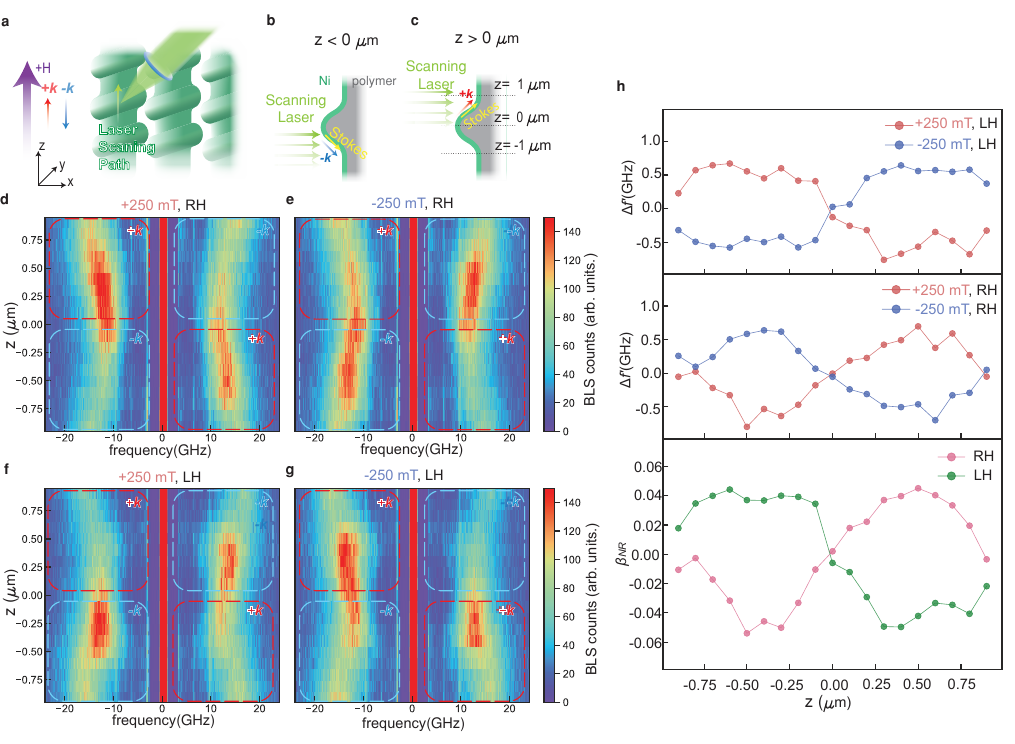}
\caption{\textbf{Nonreciprocal magnon spectra detected  along right- and left-handed ACMs via Brillouin light scattering spectroscopy.} \textbf{a}, Sketch of micro-focus BLS spectroscopy performed along the long axis of an RH-ACM. The black arrow on the ACM indicates the scanning direction in a field $H$ applied the long axis. Light scattering geometry at positions \textbf{b} $z<0~\SI{}\um$ and \textbf{c} $z>0~\SI{}\um$ leading to transferred magnon wave vectors with oppositely directed propagation directions as sketched for the Stokes signal. Color-coded spectra taken along an RH-ACM in a field of \textbf{d} +250 mT and \textbf{e} -250 mT. Color-coded spectra taken along a LH-ACM in a field of \textbf{f} +250 mT and \textbf{g} -250 mT. \textbf{h} Quantitative analysis of the nonreciprocity in terms of frequency shift $\Delta f$ for the LH-ACM (top) and the RH-ACM (center) extracted from spectra shown on the left. The bottom panel shows the magnitude of $f_{\rm NR}(\mathbf H)$ for the RH-ACM (pink) and LH-ACM (green) as a function of position $z$.
 }\label{fig2}
\end{figure}
We investigated the spin dynamics of ACMs by $\mu$-BLS at room temperature (Fig.~\ref{fig2}a). The scattering process between photons and magnons in an opaque thin film obeys the momentum conservation law for the in-plane momenta. In the BLS spectrum, the Stokes (anti-Stokes) peak corresponds to the creation (annihilation) of magnons. In the $\mu$-BLS setup, the focused laser light offers a cone of incidence angles around the optical axis of the lens. The backscattered light hence contains photons which interact with magnons exhibiting a specific distribution of in-plane wave vectors leading to the inhomogeneous broadening of BLS peaks. Assuming isotropic scattering processes, a frequency asymmetry (nonreciprocity) of the magnon band structure with respect to wave vectors $k$ parallel to the film plane is not resolved. This is no longer true if the laser light is focused on a curved surface whose normal direction does not coincide with the optical axis of the lens. In Fig.~\ref{fig2}b and c we depict the scattering configurations near the helix of an ACM. At $z<0~\SI{}\um$ and $z>0~\SI{}\um$ the Stokes (anti-Stokes) signal mainly comes from magnons propagating in parallel with (anti-parallel to) the momentum of the incident light projected into the film, i.e., the resulting magnon wave vectors $k$ are in $-z$- and $+z$-direction, respectively ($+z$- and $-z$-direction, respectively), depending on the laser spot position. The resulting imbalance in BLS spectra allows one to qualitatively analyze the frequency nonreciprocity of magnon band structures and propagation directions $-k$ versus $+k$, respectively, along the $z$-axis. 

In Fig.~\ref{fig1}c, we show spectra taken on an RH-ACM (pink) and LH-ACM (green) with a field of +250 mT applied in $z$-direction. Strikingly, the height difference between Stokes and anti-Stokes peaks is reversed for the two samples. Consequently, the sign of intensity asymmetry parameter $I_{\rm NR}$ depends on the geometrical handedness of the ACMs only as their Ni shells were deposited in the same ALD process. Also the asymmetry in frequency $\Delta f$ (Fig.~\ref{fig1}c) is inverted when changing the handedness. We attribute the handedness-dependent signal asymmetries to magnetochiral properties and nonreciprocal magnon characteristics in the ferromagnetic shell. 
The nonreciprocal characteristics introduced by geometry are further verified by spatially resolved measurements presented in Fig. \ref{fig2}d to g. The magnon spectra were taken at different positions along the long axis of the tubular shell. The position $z=0$ is put near the topmost point on the surface helix. The $z$-position-dependent spectra show clear asymmetries with respect to both $z=0$ and $f=0$ in signal strength and frequencies of signal maxima. As a function of position, both the frequency of the signal maximum and its intensity vary characteristically for the Stokes (left) and anti-Stokes (right) spectra. The observed asymmetries are reversed by either an opposite field direction (Fig. \ref{fig2}d,f versus Fig. \ref{fig2}e,g) or opposite handedness (Fig. \ref{fig2}d,e versus Fig. \ref{fig2}f,g). In the central position $z=0$ the spectra show BLS peaks with local minima of the resonance frequency. At this position, the asymmetric intensity distribution and frequency difference between Stokes and anti-Stokes signals change sign. 
 
To quantitatively analyse the asymmetries in frequency as a function of spot position (Fig. \ref{fig2}h), we apply a Lorentzian fit to the peak of each Stokes and anti-Stokes spectrum according to
\begin{equation}
{I_{\rm total}=I_0+\frac{I}{(f-f_{\rm c})^2 / \Gamma_f^2+1}},
\end{equation}
where $\Gamma_f$ is the half width at half maximum (HWHM), $f_{\rm c}$ is the central frequency, $I$ is the signal strength at the maximum, and $I_0$ is the background signal in the spectrum. We then calculate $\Delta f=\left|f_{\rm c,\text {Stokes }}\right|-\left|f_{c,\text { anti-Stokes }}\right|$
and $\Delta I=I_{\text {Stokes }}-I_{\text {anti-Stokes}}$. However, a systematic frequency shift $f^{\prime}$, independent on $z$, is inevitably present due to a slight misalignment of the laser scanning path with respect to the center of the magnetic tubes. In order to compensate for this systematic offset, we introduce $\Delta f^{\prime}=\Delta f-f^{\prime}$ and $\Delta I^{\prime}=\Delta I-I^{\prime}$, representing the position-dependent component of $\Delta f$ and $\Delta I$. Using the extracted parameters $\Delta  f^{\prime}$ and $\Delta I^{\prime}$, we calculate the asymmetry parameters as a function of $z$-position. The parameters $\Delta I^{\prime}$ for RH- and LH-ACMs at $\pm$ 250 mT (red and blue symbols) are summarized in Fig. S5, while values $\Delta f$ are summarized in the two top panels of Fig.~\ref{fig2}h. The parameters $\Delta f$ and $\Delta I$ for RH- and LH-ACMs at $\pm$ 250 mT are summarized in Fig. S8. The maximum asymmetry amounts to 0.67 GHz (0.8 GHz) in case of the LH-ACM (RH-ACM).
Considering $\Delta  f^{\prime}=\left(2 \gamma / \pi M_S\right) D k$ ~\cite{Di2015,Nembach2015}, and taking the maximum possible wave vector value $k=4\pi/\lambda_{laser}$ offered by the focussed laser light, we estimate an extrinsic DMI constant $D^{\prime}>0.112~\rm mJ/m^2$($D^{\prime}<-0.134~\rm mJ/m^2$) for LH-ACM and RH-ACM, respectively. The extracted $D^{\prime}$ values are comparable to interfacial DMI constants reported for magnetic multilayers incorporating a heavy metal~\cite{Kuepferling2023}. The latter component introduces relativistic spin-orbit interaction and enlarges magnetic damping in ferromagnetic metals. The curvature-induced ACM reported here avoids the detrimental effect of the heavy metal.

%D=(0.8*3.14*4.9*100000)/(2*176*1.1*23.684)%

To analyze the relative strength of the magnetochiral effect in ACMs and compare them to natural materials, we evaluate the relative magnitude of frequency nonreciprocity according to 
\begin{equation}
{\beta_{\rm NR}(\mathbf H)=\frac{\Delta  f^{\prime}(\mathbf{+H})}{f_{\rm S}(\mathbf{+H})+f_{\rm AS}(\mathbf{+H})}
-\frac{\Delta  f^{\prime}(\mathbf{-H})}{f_{\rm S}(\mathbf{-H})+f_{\rm AS}(\mathbf{-H})}}
\end{equation}
where $f_{\rm S}$ refers to $\left|f_c,{\text {Stokes}}\right|$ and $f_{\rm AS}$ refers to  $\left|f_c,{\text {anti-Stokes }}\right|$.
We note that $\beta_{\rm NR}$ is nominally equivalent to $g_{\mathrm{MCh}}$ as defined in Refs.~\cite{Nomura2019,Nomura2023}. In the bottom panel of Fig.~\ref{fig2}h we display $\beta_{\rm NR}$ as a function of position for the RH-ACM (pink) and the LH-ACM (green). Each dataset is anti-symmetric with respect to $z=0$. The consistent anti-phase variations observed for all three different asymmetry parameters as a function of BLS spot position are a clear signature of chirality effects controlled by geometrical handedness and magnetic field direction. The maximum peak-to-peak difference amounts to $5.4\times 10^{-2}$ at $z=-0.5~\SI{}{\um}$ for RH-ACM. This value is more than three orders of magnitude larger than $g_{\mathrm{MCh}}$ reported for a natural chiral material ($g_{\mathrm{MCh}}\approx 40\times 10^{-6}$) in Ref. \cite{Nomura2019, Nomura2023}. 

\section{Helical magnon channel around an ACM}\label{sec3}
In the following we interpret the observed frequency variation in Fig. \ref{fig2}d to g. For a polycrystalline thin film of Ni one expects a magnetic hard axis along the surface normal. Along the scanned path, the hard axis of the ACM shells gradually rotates from the $-z$-direction via the $-y$-direction toward the $+z$-direction (Fig.~\ref{fig2}a). 
%%Considering a magnetic field of 250 mT applied collinear with the y-axis, the LLG equation would suggest the largest effective field for the central position of the scanned path in Fig. 2b and 2c. At this position, the applied field is tangential to the curved Ni film and the local hard axis is perpendicular to the field. Here, one would expect a BLS spectrum with a local maximum in the detected eigenfrequency. However, the opposite behavior is experimentally observed, i.e., a local minimum in the peak frequency. One would expect the maximum. The counterintuitive frequency variation has stimulated micromagnetic simulations which we present in the following.    
Considering a magnetic field of 250 mT applied collinear with the $z$-axis, the LLG equation assumes the largest effective field in the straight part of the tube (Fig.~\ref{fig2}a) and a smaller effective field in the topmost point of the surface helix. At both of these positions, the applied field is tangential to the Ni film and the local hard axis is perpendicular to the field,  Consistently, the BLS spectra shown in Fig. \ref{fig2}d to g show local maxima (minima) of the eigenfrequencies detected within the tube (helix) regions. To further explore the microscopic nature of the low-frequency BLS signals near $z=0$ we present further experimental BLS spectra and micromagnetic simulations in the following.

\begin{figure}[h]%
\centering
\includegraphics[width=0.65\textwidth]{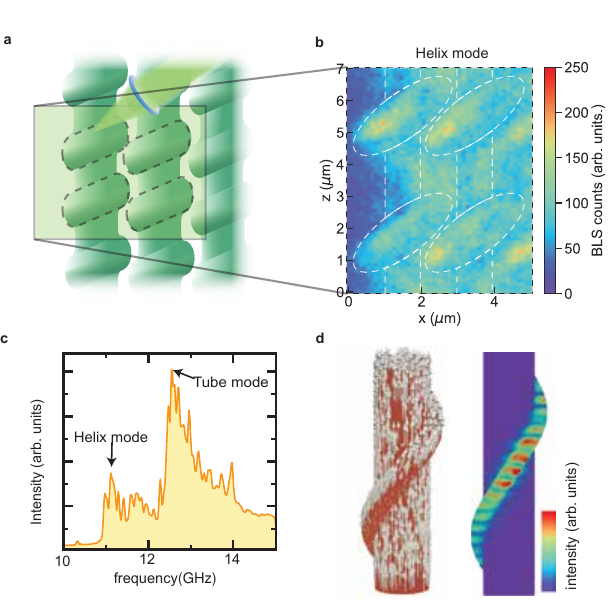}
\caption{\textbf{Helical magnon channel around an ACM.} \textbf{a,} Sketch of the measurement configuration. \textbf{b,} Spatial map of BLS signal taken at -15 GHz (Stokes signal) while +250 mT is applied to the RH-ACMs. The dashed red ovals indicate the most elevated regions due to the helix on the surface. \textbf{c,} Simulated spectrum at +250 mT for an RH-ACM with the spin structure shown in \textbf{d}. \textbf{d,} Simulated spin precession occurring along the helix at 11.12 GHz (helix mode). 
}\label{fig3}
\end{figure}

We scanned the BLS laser focus over three neighboring devices as sketched in Fig. \ref{fig3}a. The spatially resolved BLS signal (magnon distribution) was obtained by moving the sample in the $xz$-plane with 100 nm steps along $x-$ and $z-$directions while keeping a field of +250 mT along the $z$-direction and a fixed BLS detection frequency of -15 GHz (Fig.~\ref{fig3}b). Further data taken at $\pm 13$ and $\pm 17~$GHz are displayed in Supplementary Fig. S6. The BLS mapping of Fig.~\ref{fig3}b shows a spatial intensity distribution which exhibits a weak signal strength in the tube region and a large one in the helix region. The largest signals are found in the elevated helical reliefs indicated by the dashed ovals. Within an oval, the signal distribution at -15 GHz is inhomogeneous suggesting a nonreciprocal character of magnons along the helical surface relief. The signals at $\pm 13~$GHz are strong in the ovals as well. However, at $\pm 17~$GHz the large signals occur in the tube region (compare Supplementary Fig.~S6).

Using the GPU-accelerated software MuMax3~\cite{vansteenkiste_design_2014} we simulated a Ni ACM consisting of a tube with inner radius of 220 nm and thickness of 30 nm which intersects a hollow helix with an ellipsoidal cross-section. The helix had a pitch of 2000 nm, diameter of 740 nm, cross-sectional inner major and minor radii of 120 nm and 70 nm, respectively, and a thickness of 30 nm, as detailed in the Methods section. The diameter is smaller compared to the real sample as a consequence of the limited computational power. An external field of $+250$ mT was applied along the $z$-axis. To investigate the dynamic response a small dynamic field $h=h_0 \operatorname{sinc}\left(2 \pi f_{\rm c}\left(t_{-} t_0\right)\right)\hat{x}$ with pulse amplitude $\mu_0h_{0} = 3~{\rm mT}$, cut-off frequency $f_{\rm c}=15~\rm GHz$, time $t$ and time-offset $t_{0}=26.7~{\rm ns}$ was applied to a 20 nm wide strip forming a ring-like excitation region around the center. Six repetitions of periodic boundaries (PBCs) were applied along the $z$-axis to prevent finite-size induced vortex states at the ends.
Following the procedures outlined in the Methods, we obtain a spectrum as presented in Fig.~\ref{fig3}c when considering the spin configuration of an RH-ACM shown in Fig.~\ref{fig3}d (left).
Multiple peaks are resolved in Fig.~\ref{fig3}c which appear mainly in two groups. Coming from low frequencies, one group exists up to about 12.3 GHz. The maximum signal strength shows up at 12.56 GHz followed by a second group of peaks extending to larger frequencies. We are particularly interested in the low-frequency regime and display the spatial distribution of the magnitude of the complex dynamic amplitude at 11.14 GHz in Fig.~\ref{fig3}d (right). The spin-precessional motion of this low-frequency excitation occurs along the helical relief on the surface. The spin-precessional amplitude is distributed asymmetrically along the surface helix. The tube region is not excited. The simulations are qualitatively consistent with the experimental observation presented above and in Supplementary Fig. S6. The remaining discrepancy concerning eigenfrequencies is attributed mainly to the different diameters.

%By applying +50 mT to LH ACMs, we selectively collect signal at -9 GHz (+ 9 GHz) channel in Stokes ( anti-Stokes) spectra, mapping out spatial distribution in 5 X 7 ($\mu$m$^{2}$) area (Fig. 4b) and presenting the results in Fig. 4d and Fig.4f. The scanning area covers 4 helices of two parallel align LH ACMs, to ensure reliability of the experiment. As shown in Fig. 4d, -9 GHz signal are mainly comes from the right side of the helix while +9 GHz signal from left side ( Fig. 4f). Since the Stokes signal at right side (left side) originates from the scattering process with +k (-k) propagating magnon, the asymmetry distribution, hence, uncovers nonreciprocal spin dynamics in space, evident that magnons on LH ACMs prefers to propagate along +k direction. This simultaneously confirmed by the opposite distribution of anti-Stokes signal. 

\section{Nonreciprocity at remanence induced by field-reprogrammable toroidal moment vectors}\label{sec4}

\begin{figure}[h]%
\centering
\includegraphics[width=1\textwidth]{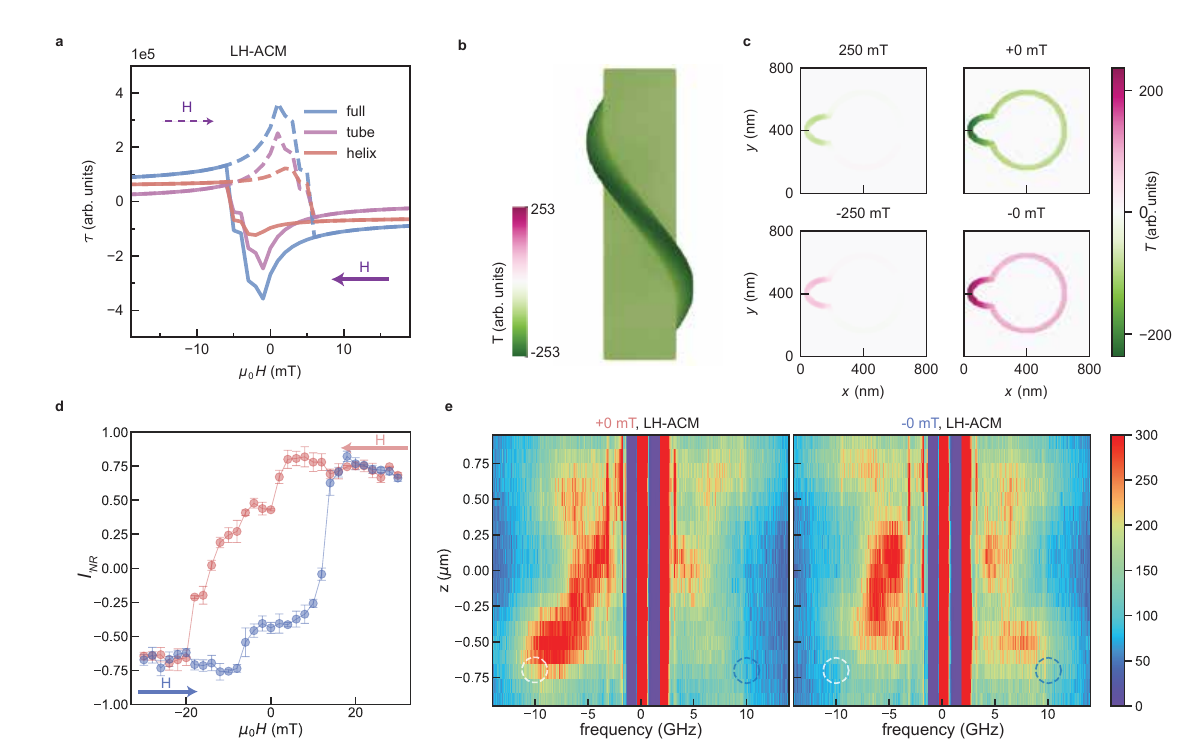}
\caption{\textbf{Programmable toroidal moments and nonreciprocal transport at remanence. } \textbf{a,} 
Toroidal moment $\tau$ of a LH-ACM (blue curve) simulated as a function of axial field $H$ and its sweep direction indicated by yellow arrows. The magenta and orange curves decompose the toroidal moment of the LH-ACM in the contributions from the tube and helix regions, respectively. The field was swept from negative to positive (dashed lines) and back (solid lines). The hysteretic behavior enables a reprogrammable $\boldsymbol{\tau}$ at $H=0$.  \textbf{b,} Simulated density distribution of $\tau$ of a LH-ACM at remanence state ($+0$~mT) following saturation at $+250$~mT. \textbf{c,}  Simulated cross-section of $\tau$ distribution on a LH-ACM when varying the magnetic field from +250 mT, to +0 mT, $-$250 mT, and $-$0 mT. The cross-section was taken at the central plane of the ACM. \textbf{d,} Intensity nonreciprocity $I_{\rm NR}$ as a function of magnetic field and sweep direction, showing a hysteretic behavior at position $z= -0.7$ \SI{}{\um}. \textbf{e,} Color-coded magnon spectra $+0$~mT and $-0$~mT obtained along the central axis along the top surface of a LH-ACM at $\pm$0 mT after applying $+50$~mT and $-50$~mT, respectively. The white (blue) circles indicate the same positions on the $z$-axis and highlight clearly asymmetric Stokes (anti-Stokes) signals which substantiate their dependence on the magnetic history.
}\label{fig4}
\end{figure}

The data of Fig.~\ref{fig2}d to g showed that for a given handedness the nonreciprocity of BLS signals depended on the field orientation. Considering Eq. \ref{eq:nonrec_toroidal}, we expect a sign change of the toroidal moment vector with positive and negative $H$. Figure \ref{fig4}a to c summarizes the simulated magnetic field dependence of $\boldsymbol{\tau}$ in an LH-ACM. In Fig.~\ref{fig4}a, we show computed toroidal moment values according to Eq. \ref{eq:tormom} when sweeping the magnetic field from negative (dashed curves) to positive direction and back (solid curves). We note that the value of the toroidal moment vector depends on the choice of origin. As long as this choice is made consistently, the change in toroidal moment is independent of the choice of origin~\cite{Spaldin08}. We have taken the center of the tube as the origin. 
In the supplementary material, we show that simulations performed for the RH-ACM result in a mirror image of Fig. \ref{fig4}a, which reflects the opposite handedness of the ACMs.

In Fig. \ref{fig4}a we decompose $\tau_{\rm full}$ into a contribution from the tube region $\tau_{\rm tube}$ (purple curves) and a contribution from the helix region $\tau_{\rm helix}$ (red curves). As shown in Fig.~\ref{fig4}a, $\tau_{\rm tube}$ vanishes when the magnetic field increases (decreases) to +10 mT (-10 mT), whereas $\tau_{\rm helix}$ remains at a finite value in the presence of a strong magnetic field ($\pm$ 250 mT). Besides, at the remanent state ($\pm$ 0 mT), the LH-ACM sustains a substantial amplitude of $\tau_{\rm full}$, which is further visualized in Fig.~\ref{fig4}b where the full toroidal moment density $\boldsymbol{T}$ is shown for +0 mT. Note that in Fig.~\ref{fig4}a, $\tau_{\rm tube}$ appears to have a larger contribution to $\tau_{\rm full}$, which is attributed to the larger volume of the tube region. Figure~\ref{fig4}c shows cross-sections of the toroidal moment density $\boldsymbol{T}$ of the LH-ACM for a sweep from -250 mT to $-$0 mT and a sweep from 250 mT to 0 mT. In both cases, the toroidal moment is large in remanence, with an appreciable contribution originating from the helix region.

In Fig. \ref{fig4}d, we report BLS spectra taken on one and the same LH-ACM after saturating it with a magnetic field along the $+z$-($-z$-)direction expecting $-\tau(+\tau)$ from Fig.~\ref{fig4}c. The BLS spectra were obtained in zero magnetic field after saturating at +50 mT (labeled +0 mT) and $-$50 mT (labeled $-$0 mT). Comparing the BLS signals near the dashed circles in Fig.~\ref{fig4}d, the Stokes peak of +0 mT is located at higher frequency with stronger intensity than in the $-$0 mT spectra, while anti-Stokes peaks show the opposite behavior. The spectra exhibit position-dependent nonreciprocity consistent with the +250 mT spectra of Fig.~\ref{fig2}d to g. The corresponding spectra on RH-ACM are given in Fig.~S3, which shows opposite position dependence with respect to Fig. \ref{fig4}d. In Fig. \ref{fig4}e we display the full hysteresis curve of the evaluated intensity asymmetry $I_{\rm NR}$. Here, we swept the magnetic field from +30 mT to $-$30 mT and then back to +30 mT while acquiring BLS intensities at frequencies $+10$ GHz and $-10$ GHz at the $z$ coordinate highlighted by the dashed circles. $I_{\rm NR}$ was extracted via \begin{equation}
{I_{\mathrm{NR}}(f_0, \mathbf H)=\frac{I_{\text {S}}(f_0,\mathbf H)-I_{\text {AS } }(f_0, \mathbf H)}{(I_{\text {S }}(f,\mathbf H)+I_{\text {AS } }(f_0, \mathbf H))/2}}.
\end{equation} These independent data substantiate the non-vanishing chirality of an ACM at zero magnetic field and the field-programmable nonreciprocity. The unique surface topography of ACMs favors a spirally aligned spin texture which consequently preserves the chirality with $\tau\neq0$ at zero magnetic field. 
To explore the reproducibility of the chirality memory effect in an ACM, we sequentially applied $\pm$30 mT, and extracted $I_{\mathrm{NR}}$ at $\pm 0$~mT for five times before applying the opposite magnetic field history. The data are displayed in Fig. \ref{fig1}d and Fig.~S4c,d. Within the noise level, $I_{\rm NR}$ exhibits only two levels reflecting two different magnon reciprocities reproducibly induced at zero field. The hysteretic curves of the LH- and RH-ACM are acquired at position $z= -0.7 \SI{}\um$, showing opposite nonreciprocity in Fig. S4a and S4b.

%\section{Geometrical tunning of magnon nonreciprocity in ACMs}\label{simulation}
\section{Optimization of magnon nonreciprocity in ACMs}\label{simulation}
\begin{figure}[h]%
\centering
\includegraphics[width=0.9\textwidth]{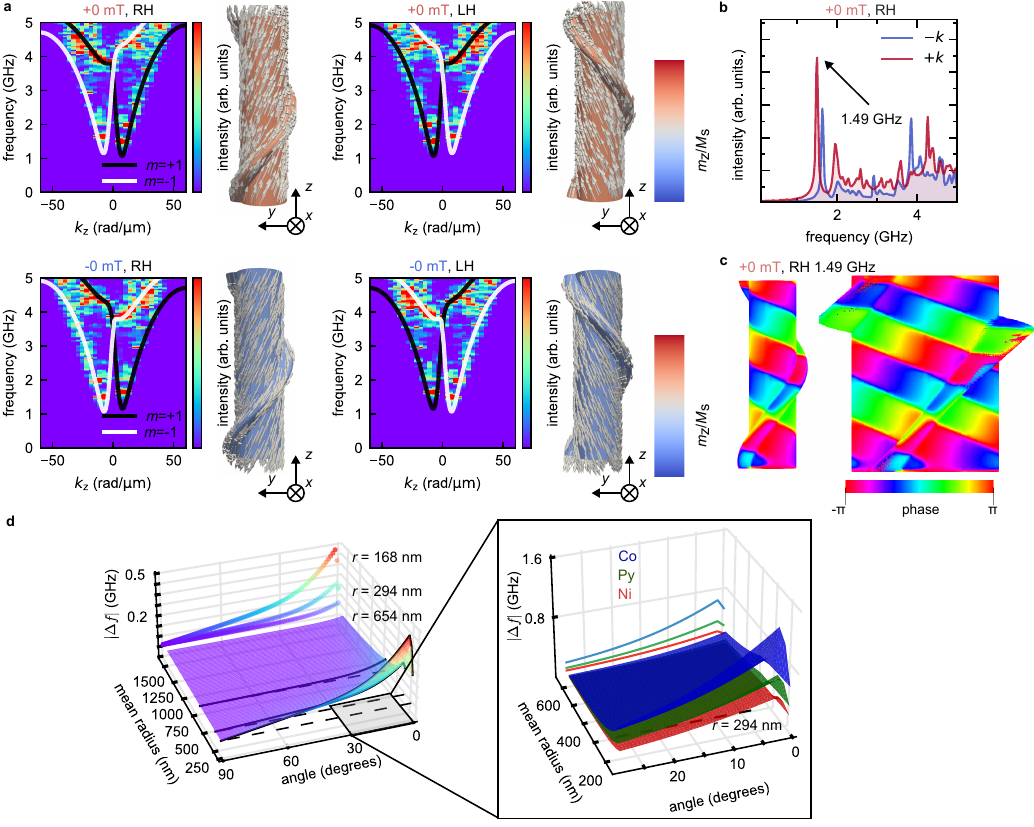}
\caption{\textbf{Optimization of reprogrammable nonreciprocity in ACMs.} \textbf{a,} Numerical simulation results of spin wave dispersion relations $f(k_{\rm z})$ on RH and LH Ni tubes at $\pm 0$~mT. The 3D illustrations provide the remanent magnetic textures leading to $f(k_{\rm z})$ shown on their left sides. \textbf{b,} Contributions of $\pm k$ magnons to the spectra of a RH-ACM at +0 mT. \textbf{c,} Phase distribution of magnon modes at 1.49 GHz on an ACM. \textbf{d,} Optimization of nonreciprocity ($\Delta f$) according to Eq. \ref{eq:disp} by varying geometrical and materials parameters as well as the angle the helical spins take with respect to the longitudinal axis.
}\label{fig5}
\end{figure}

In the following we discuss how to engineer and optimize nonreciprocity in ACMs. To this end, first, the dispersion relation of magnons $f(k_{\rm z})$ on the tubular segment is obtained via dynamic micromagnetic simulations. Both right-handed and left-handed ACMs are considered. Figure~\ref{fig5}a shows the numerical results as color-coded graphs obtained for four configurations provided by differently oriented spin textures and the different handedness (shown on the right of each graph). Thereby, we realize four combinations of helicity of spins $\chi=\pm1$ and polarization $p=\pm1$. $p$ is evaluated from $\boldsymbol{m}_0$ projected on the $z$-axis (for details, see Supplymentary materials.
The top row of Fig.~\ref{fig5}a shows the dispersion obtained for an RH- and LH-ACM at zero field, where the system was first saturated along the positive $z$-direction. Due to their opposite handedness, the spin textures of the two tubes have identical polarity but opposite helicity (handedness). The color intensity map shows the resulting eigenfrequencies of the tube segments. The noisy appearance is due to the finite frequency and field resolution of the performance-hungry simulations on ACMs. The numerical results demonstrate a clear frequency nonreciprocity between $\pm k_z$, whose sign is opposite for RH- and LH-ACMs. We compare the numerical results obtained on ACMs with dispersion relations calculated analytically by an expression provided for nanotubes with conical spin texture \cite{salazar-cardona_nonreciprocity_2021} (see Methods). The analytical results (solid lines) agree almost quantitatively with the branches resolved by the full-fledged numerical simulations applied to ACMs, when we consider an effective radius $r_{\rm st}$ of 300 nm for the straight (st) nanotube. The two minima shown in the numerical dispersion can be identified with the two azimuthally propagating modes $m=\pm1$, with $m$ being the azimuthal index. The frequency nonreciprocity is opposite for RH- and LH-ACMs highlighting the geometrical programmability of magnon transport via the surface helix. Stimulated by ~\cite{DelSer2021} and considering their peculiar geometry and the unidirectional magnon flow along the $z$-axis we call the created ACMs magnonic Archimedean screws. \\
Furthermore, we show that the nonreciprocity at remanence is field tunable. After first saturating the system with a field applied along the $-z$-direction and then approaching 0 mT, we reverse the nonreciprocal dispersions for the RH- and LH-ACMs (Fig.~\ref{fig5}a, bottom row). The different contributions of $+k_z$ and $-k_z$ magnons can be clearly shown by integrating the intensity of the dynamic magnetization (Fig. \ref{fig5}a) over the positive and negative wavevectors, respectively. The resulting spectrum of the RH-ACM is shown in Fig.~\ref{fig5}b and clearly demonstrates that $+k_z$ magnons dominate the absolute frequency minimum in the dispersion. In Fig. \ref{fig5}c we display the phase profile of the mode with the frequency of the main peak at 1.49 GHz. The magnons are seen to propagate in a helical manner along the tube. Unwrapping the ACM onto a plane reveals a phase jump induced by the surface helix (Fig.~\ref{fig5}c).\\
There are two crucial parameters responsible for the nonreciprocity that are related to the geometry of the ACM: one is the angle $\eta$ of the magnetization $\boldsymbol{m}_0$ with respect to the longitudinal tube axis, which is related to the surface helix angle, and the other one is the radius $r$ of the tube region. Using the theoretical formalism introduced by Salazar-Cardona et al.\cite{salazar-cardona_nonreciprocity_2021}, we studied the parameter-dependent nonreciprocity of Ni ACMs in greater detail. For this, we assumed that the opening angle $\alpha$ of the conical spin structure relates to $\eta$ and the straight tube radius $r_{\rm st}$ to $r$. In Fig.~\ref{fig5}d, the computed nonreciprocity $\Delta f$ for long tubes is shown as a function of angle $\alpha$ and radius $r_{\rm st}$. The nonreciprocity increases for both smaller radius and smaller angles. We show line cuts for three radii: $r=168,~294,~{\rm and} ~ 654$ nm. Here, 294 nm is the effective radius of the tube, whereas the radius of 654 nm is comparable to the experimentally realized ACM presented in this work. We do not display the parameter regime of very small values in which the hierarchy of azimuthal modes $m$ is reversed as TPL sets a lower limit to $r_{\rm st}$. The minimum radius reached by state-of-the-art TPL amounts to 50 nm. The non-reciprocity can be further enhanced by the choice of polycrystalline ferromagnetic material. In Fig.~\ref{fig5}d the close-up shows a comparison of the nonreciprocity for three ferromagnets: Ni, Py, Co (Methods). Both Py and Co have a larger nonreciprocity than Ni. In particular, Py shells are also accessible via ALD ~\cite{Giordano2021,Giordano2023a}.

\section{Conclusion}\label{sec6}
We reported the successful creation of artificial chiral magnets which support nonreciprocal magnon transport at zero magnetic field. Our work unveils the interplay between geometrical handedness and nonreciprocity. It demonstrates that chirality can be imprinted at room temperature by surface engineering of a conventional ferromagnet. Magnons on LH- and RH-ACMs show opposite nonreciprocities in both magnon frequencies and intensities. The nonreciprocity can be reversed by reversing the magnetic field direction. The artificial magnetic chirality manifests itself in a pronounced magnon nonreciprocity chiral parameter $\beta \rm_{NR} \rm = 5.4\times 10^{-2}$, which are three orders of magnitude higher than the reported value in bulk chiral magnets~\cite{Nomura2023}.

The presented ACMs are created by combining TPL and ALD, which is compatible with mass production of components in microelectronics. In particular, the ACMs have been realized based on a polycrystalline magnetic material. Such a material is preferred in technologies over single crystals which were explored in the case of natural chiral magnets. The chirality of ACMs originates from the geometrical handedness and a helical Ni surface on a central cylinder. This structure gives rise to a magnonic Archimedean screw. The significant magnon nonreciprocity of our 3D nanomagnetic elements in the GHz frequency regime fuels the prospects of 3D interconnected magnonic circuits with uni-directional signal transmission in vertical directions. Such elements enhance the integration density and energy efficiency of magnonics.

%Schematics of the spatial resolved BLS measurement on ACMs.  \textbf{c,d}, spectra on RH ACMs under +250 mT and -250 mT magnetic field while varying the position along ACMs. \textbf{e, f}, spectra on LH ACMs. \textbf{g, h},  frequency difference ($\Delta$f) of Stokes peak and anti-Stokes peak of the magnon spectra on LH (red dots), RH (blue dots). \textbf{i,j}, signal intensity difference ($\Delta$A) of Stoke peak and anti-Stoke peak of the magnon spectra on LH (red dots), RH (blue dots). \textbf{k, l}, static magnetization of a chiral tubes of a RH, LH ACMs. \textbf{m, n,} Distribution of toroidal moment density on RH, LH ACMs. 

%density distributions of toroidal moment when scanning magnetic field from +250 mT, to +0 mT, -2 mT, -4 mT, -6 mT and -250 mT. \textbf{c, d,} toroidal moments of g-i Zero magnetic field ( +0 mT, $-$0 mT ) BLS spectra on LH ACMs after saturated at 50mT and $-$50 mT \textbf{a} simulation results of the magnetic hysteresis loop when applying magnetic field along z axis, parallel to tubes. \textbf{b} variation of toroidal moment while magnetic field is scanning from +250 to +0mT, and to -250 mT.  \textbf{b }static magnetization of a chiral tube at 0 mT when sweeping magnetic field from positive to negative field. \textbf{c,} from negative to positive. Numerical simulation result of spin waves dispersions on chiral tubes at 0 mT, for tubes which was saturated at positive field, +0 mT (a), and negative field, $-$0 mT (b). c) the difference of the +0 mT dispersion and $-$0 mT dispersion.

\clearpage

%\textbf{a,} simulation results of the magnetic hysteresis loop when applying magnetic field along z axis, parallel to tubes. \textbf{b }static magnetization of a chiral tube at 0 mT when sweeping magnetic field from positive to negative field. \textbf{c,} from negative to positive. Numerical simulation result of spin waves dispersions on chiral tubes at 0 mT, for tubes which was saturated at positive field, +0 mT (a), and negative field, $-$0 mT (b). c) the difference of the +0 mT dispersion and $-$0 mT dispersion.

\clearpage
\bibliography{ACM.bib}

\clearpage
%% \bibliography{bib}

%% if required, the content of .bbl file can be included here once bbl is generated
%%\input sn-article.bbl
%\bibliographystyle{Science}

\section{Methods}\label{sec6}

\subsection{Sample Preparation}\label{subsec2}
%\textbf{IDENTICAL TO SECTION 2}
The magnetic chiral tubes were fabricated by combining two-photon lithography (TPL) and atomic layer deposition (ALD). We applied the additive manufacturing methodology described in \cite{Guo2023a} to 3D polymer wires which contained helical reliefs. They were prepared by TPL using a Photonic Professional GT+ system (Nanoscribe Inc., Germany) in three steps. Firstly, negative photoresist IP-Dip was dropped onto a fused-silica substrate (25 x 25 $\mathrm{mm^2}$ with a thickness of 0.7 mm). Secondly, an infrared femtosecond laser (wavelength: 780 nm, power: 20 mW) was focused inside the resist exploiting the dip-in laser lithography (DILL) configuration for the exposure. Thirdly, the whole substrate was immersed in propylene glycol monomethyl ether acetate (PGMEA) for 20 min and isopropyl alcohol (IPA) for another 5 min. After the polymer was dried in ambient conditions, the sample was put into a hot wall Beneq TFS200 ALD system. We conformally coated the polymer with a 30-nm-thick Ni shell after depositing 5-nm-thick Al$_{2}$O$_{3}$ using the plasma-enhanced ALD process presented in ~\cite{Giordano2020}. The detailed preparation process is presented in (Fig. S1). 
\subsection{Brillouin light scattering microscopy}\label{subsubsec2}
The spin dynamics were investigated by $\mu$-BLS at room temperature (Fig. S2). The samples were mounted on a piezo stage by which the sample was moved in steps of 50 nm underneath the laser focus. Positive and negative external magnetic fields were applied by permanent magnets mounted in different orientations along the $x$-axis. The ACMs were positioned parallel to the $x$-axis. The green laser (wavelength: 532 nm) with a power of 3 mW was focused on the surface of the helical magnet using a 100X objective lens with a numerical aperture of 0.75. The s-polarized component of the scattered light was passed through a Glan-Taylor polarizer and sent to a six-pass tandem Fabry-Pérot interferometer.  In the $\mu$-BLS setup, the focussed laser light offers a cone of incidence angles around the optical axis of the lens. The backscattered light contains photons which interact with magnons with different in-plane wave vectors $+k$ and $-k$. The magnitudes of $k$ vectors ranges from 0 to about $17.7~\mathrm{rad}/\mu$m.

\subsection{Simulation}\label{subsubsec2}

Micromagnetic simulations were conducted using the MuMax3 software \cite{vansteenkiste_design_2014} which solves the Landau-Lifshitz-Gilbert (LLG) equation on a finite difference grid. We considered a Ni ACM consisting of a tube with inner radius 220 nm and thickness 30 nm which intersects a hollow helix with ellipsoidal cross-section. The helix had a pitch of 2000 nm, diameter of 740 nm, cross-sectional inner major and minor radii of 120 nm and 70 nm respectively, and thickness 30 nm. The saturation magnetization was set to $M_{\rm s}=490~\mathrm{kA/m}$ and the exchange stiffness to $A_{\rm ex}=8~\rm{pJ/m}$. The system was discretized into 160×160×384 cells of dimension $5\times5\times5.2~\rm{nm^3}$. Six repetitions of periodic boundary conditions along the $z$-direction were used.

Hysteresis diagrams of the structures were computed by sweeping an applied field parallel to the tube axis with a $2^{\circ}$ misalignment between $+1~\rm{T}$ and $–1~\rm{T}$ and back to $+1~\rm{T}$. Additionally, a constant background field of 0.7 mT along the $x,y$-diagonal was applied. The magnetic ground state was computed in between specified field increments by first using the steepest conjugate gradient method \cite{exl_labontes_2014} to minimize the energy and then solving the LLG without precessional term. The resulting ground states provided the initial state for the computation of the toroidal moment and the dynamic behavior.

The toroidal moment was computed per layer using Eq. \ref{eq:tormom} using the tube axis as the origin. The plots in Fig. \ref{fig4}c-e show the product $\mathbf{r}\times\mathbf{m}\left(\mathbf{r}\right)$ projected on the $z$-axis.

The dynamic simulations were conducted following the methodology in the main text. The dynamic field was confined to a strip of width 20 nm. The simulations were run for a total time of 53.3 ns and the magnetization was sampled on the surface of the tube along the tube axis every 33.3 ps. The damping was set to $\alpha=10^{=-3}$ and increased quadratically to 1 near the ends of the structure. For the numerical results shown in Fig. \ref{fig3}d, care had to be taken that no asymmetries due to field misalignment were present. Therefore, the ground state from the hysteresis was used as an initial state, then the background signal was removed and the applied field was fully aligned with the tube axis. The ground state was subsequently recalculated. The gilbert damping was set to $\alpha=10^{-3}$. A fast Fourier transform (FFT) was performed over the dynamic magnetization components $m_{\rm x},m_{\rm y}$. Then, the intensity of the complex dynamic magnetization $m_{\rm d}=m_{\rm x} + im_{\rm y}$ was computed. Its spatial distribution is shown in Fig.~\ref{fig5} d,e and the integrated intensity is shown in Fig.~\ref{fig5} c. Finally, the dispersion shown in Fig.~\ref{fig5} was obtained by performing a 2D FFT over the magnetization sampled on the tube along the $z$-axis.

To clearly visualize the magnon propagation in real space, simulations were performed on longer tubes (4000 nm). No periodic boundaries were applied. To minimize stray field effects at the edges, the tube radius was reduced to 125 nm and the helix end-to-end diameter was set to 200 nm. The dispersion of this structure is shown in Supplementary.

\subsection{Analytical dispersion}\label{subsubsec2}

The dispersion in Fig.~\ref{fig5} is plotted together with data obtained from the analytical model proposed by Salazar et al. \cite{salazar-cardona_nonreciprocity_2021} for nanotubes with helical equilibrium magnetization. The dispersion is given by

\begin{equation}
    \omega_{m}\left(k\right)=\omega_{M}\left[\mathcal{A}_m\left(k\right) + \sqrt{\mathcal{B}_m\left(k\right)\mathcal{C}_m\left(k\right)} \right],
    \label{eq:disp}
\end{equation}

with $\omega_M=\gamma\mu_{\rm 0}M_{\rm s}$, $\gamma$ is the gyromagnetic ratio and $k$ the wavevector. The index $m$ denotes the azimuthal mode. $\mathcal{A}_m\left(k\right), \mathcal{B}_m\left(k\right), \mathcal{C}_m\left(k\right)$ are the dynamic stiffness fields. The frequency non-reciprocity is determined by the magnetochiral stiffness field $\mathcal{A}_{k}^m=-\chi \mathcal{K}\qty(m, k)\sin \qty(\theta) + p\qty(\mathcal{N}\qty(m,k)-\frac{2m\lambda_{exc}^2}{b^2})\cos \qty(\theta)$. Here, $\theta$ is the angle of the magnetization with respect to the tube axis, $p=\pm 1$ is the polarity of the magnetization and $\chi=\pm 1$ is the helicity. We note that there is a minus sign difference for the expression of $\mathcal{A}$ \cite{korber_spin_2023}. The functions $\mathcal{K}\qty(m,k),\mathcal{N}\qty(m,k_z)$ are demagnetizing factors. The data shown in Fig.~\ref{fig5} is obtained from Eq. \ref{eq:disp} in the ultra-thin shell approximation where $t\approx\lambda_{\rm exc}\ll R$, with $t$ the thickness, $\lambda_{\rm exc}$ the exchange length and $R$ the mean radius of the tube. Complete expressions are given in the supplementary information.

\backmatter

\section*{Declarations}

\bmhead{Acknowledgments}
This research was supported by the SNSF via grant number 197360. The authors thank PD Dr. Oleksandr Dobrovolskiy, Dr. Jun He, Dr. Yang Liu for the helpful discussion. The simulations and post-processing have been performed using the facilities of the ScientificIT and Application Support Center of EPFL.

\bmhead{Availability of data and materials}
The data presented in the main text and the supplementary information are available from the corresponding authors upon reasonable request.
\bmhead{Code availability }
The code used in this study is available from the corresponding author
upon reasonable request.
\bmhead{Competing interests }
The authors declare no competing interests.
\bmhead{Author contributions}
M.X. and D.G. conceived the study and the design of the devices. M.X., A.D., and D.G. prepared figures. M.X. wrote the main manuscript text, with support from D.G. and A.D. M.X. performed the measurements. A.D. conducted micromagnetic simulations. H.G. fabricated the devices. M.X., A.D., and D.G. processed and analyzed the measurements and simulations. All authors reviewed the manuscript and provided critical feedback. D.G. supervised this project.

\clearpage

\end{document}

\maketitle

% \section*{Supplemental materials}
\clearpage

\section{ACM fabrication}\label{secS1}

\begin{figure}[h]%
\centering
\includegraphics[width=0.9\textwidth]{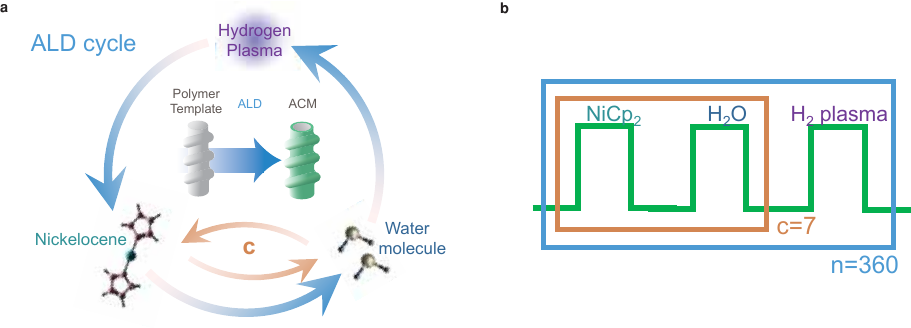}
\caption{\textbf{Plasma-Enhanced Atomic Layer deposition (ALD)} \textbf{a,} Sketch of an ALD cycle for Ni. The ALD cycle consists of c step of NiCp$_{2}$/H$_{2}$O (orange arrows), resulting in the formation of nickel oxide, and followed by a hydrogen plasma (purple), reducing nickel oxide to metallic nickel. This cycle was repeated n times as in \textbf{b}. 30 nm conformal Ni coating was achieve by $n=360$ while keeping $c=7$ at a substrate temperature of $T=180 ^\circ C$. Then an in-situ annealing process at $350 ^\circ C $ under a mixture of pure hydrogen and nitrogen flow was performed.
}
\label{S-FigALD}
\end{figure}

\section{BLS setup}\label{secS1}
\begin{figure}[h]%
\centering
\includegraphics[width=0.9\textwidth]{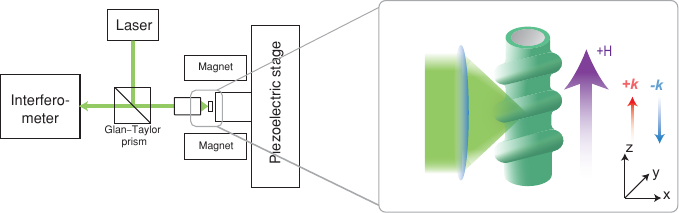}
\caption{\textbf{Illustration of a room temperature $\mu $-BLS}, the ACMs are suspended from the substrate and mounted to a piezostage. S-polarized green laser light is selected using a Glan-Taylor prism and then focused by a 100x objective onto the samples. The inelastically scattered light is collected by the same objective, and its p-polarized components are subsequently sent to interferometer.
}
\label{S-FigBLS}
\end{figure}
\clearpage

\section{Spatially resolved spectra of RH-ACM at remanence}\label{S-Fig0mTRH}

\begin{figure}[h]%
\centering
\includegraphics[width=0.9\textwidth]{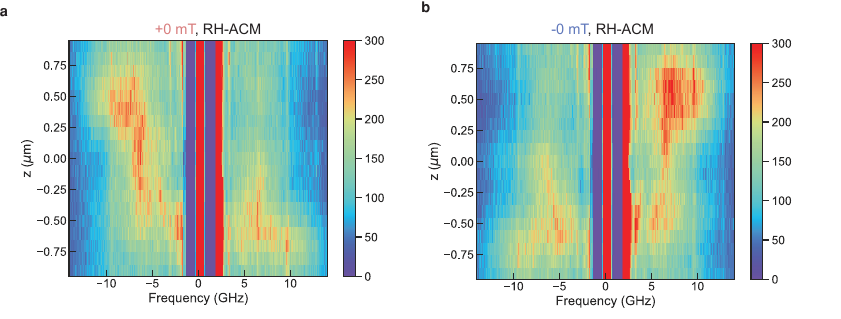}
\caption{\textbf{ Spatial scanning on RH-ACM at the remanent state.} Color-coded magnon spectra \textbf{a} +0 mT and \textbf{b} -0 mT obtained along the central axis along the top surface of a RH-ACM after applying +30 mT and $-$30 mT, respectively. }
\label{S-Fig0mTRH}
\end{figure}

In Fig. \ref{S-Fig0mTRH}, we report BLS spectra taken on one  RH-ACM after saturating it with a magnetic field along the $+z$-($-z$-) direction. The BLS spectra were obtained in zero magnetic field after saturating at +30 mT (labeled +0 mT) and $-$30 mT (labeled $-$0 mT). Fig. \ref{S-Fig0mTRH}a,b show mirrored nonreicprocity dependence, which are opposite to the corresponding dependence on RH-ACM (Fig. 4 of the main text). We note that spectra show position-dependent nonreciprocity.  
\clearpage

\section{Nonrecprocity reprogramming on RH-ACM}\label{secS1}
\begin{figure}[h]%
\centering
\includegraphics[width=1\textwidth]{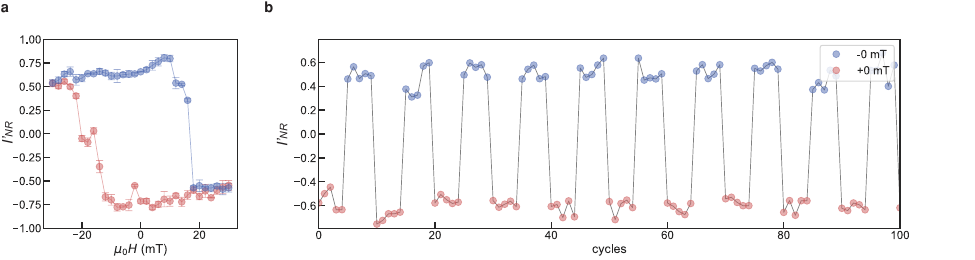}
\caption{\textbf{Nonreciprocity reprogrammed RH-ACM} \textbf{a} Intensity nonreciprocity $I_{\rm NR}$ as a function of magnetic field and sweep direction on RH-ACM at position $z= -0.7$ \SI{}{\um}. 
\textbf{b} Nonreciprocity parameter $I_{\rm NR}$ extracted from spectra taken at
$\mu_0 H=\pm 0$ mT after applying a magnetic field of $\pm 30$ mT in opposite axial directions of a RH-ACM. $I_{\rm NR}$ was measured five times (cycles) before realizing a reversed magnetic field history at position z= -0.7 \SI{}{\um}. }
\label{S-FigCode}
\end{figure}

The magnon nonreciprocity at remanence originated from the nonvanishing torodial moments at zero field, which are defined by structural handedness and magnetic field history. In Fig. \ref{S-FigCode}a, we display the hysteresis curve of the evaluated intensity asymmetry $I_{\rm NR}$ of RH-ACM. Here, we swept the magnetic field from +30 mT to $-$30 mT and then back to +30 mT while acquiring BLS counts at frequencies $+10$ GHz and $-10$ GHz. As shown in Fig.\ref{S-FigCode}b, $I_{\rm NR}$ of RH-ACM can be switched from a positive to a negative value (negative to a positive value) by saturating with a $-$30 mT (+30 mT) external magnetic field. We note that the $I_{\rm NR}$ provides qualitative information for monitoring the changing of the magnetic state, meaning that the sign reversal of $I_{\rm NR}$ corresponds to the reversal of toroidal moment but the amplitude variation does not directly reflect the modification of magnetic textures.

\clearpage
\section{Intensity nonreciprocity}\label{secS1}

\begin{figure}[h]%
\centering
\includegraphics[width=0.7\textwidth]{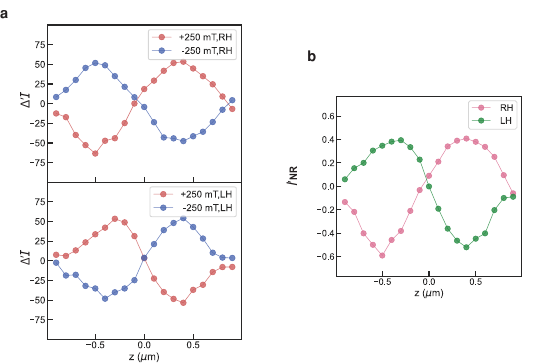}
\caption{\textbf{Intensity nonreciprocity}
\textbf{a}, Quantitative analysis of the nonreciprocity in terms of the magnitude difference of $\Delta ^{\prime}I$ for the RH-ACM (top) and the LH-ACM (bottom). The left panel \textbf{b} shows the magnitude of $I^{\prime}_{\rm NR}(\mathbf H)$ for the RH-ACM (pink) and LH-ACM (green) as a function of position $z$.}
\label{S-Fig_DeltaI}
\end{figure}
We evaluate the magnitude of intensity nonreciprocity according to
\begin{equation}
{I^\prime _{\rm NR}(\mathbf H)=\frac{\Delta I^{\prime}(\mathbf{+H})}{I_{\rm S}(\mathbf{+H})+I_{\rm AS}(\mathbf{+H})}
-\frac{\Delta I^{\prime}(\mathbf{-H})}{I_{\rm S}(\mathbf{-H})+I_{\rm AS}(\mathbf{-H})}}
\end{equation}

where $I_{\rm S}$ refers to $\left|I_{\text{Stokes}}\right|$ and $I_{\rm AS}$ refers to  $\left|I_{\text{anti-Stokes }}\right|$.
We note that $I^\prime _{\rm NR}$ represents the difference in peak amplitude while $I_{\rm NR}$ used in the main textxt corresponds to the difference at a fixed frequency.

\clearpage

\section{Magnon tube mode}\label{secS1}

\begin{figure}[h]%
\centering
\includegraphics[width=0.9\textwidth]{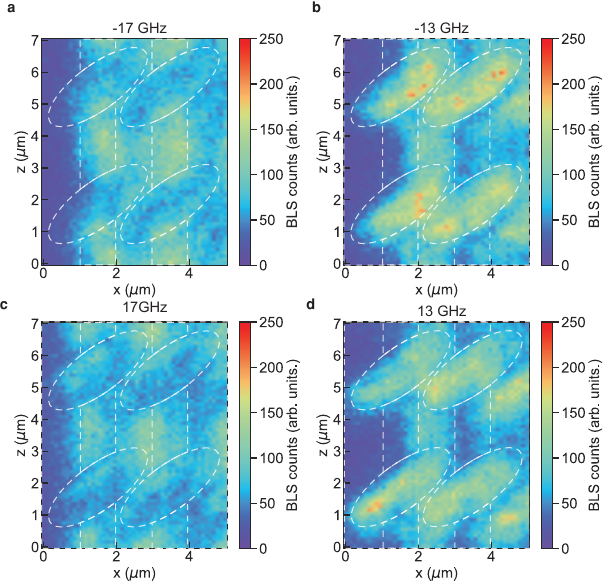}
\caption{\textbf{Magnon tube mode and helix mode.} Spatially resolved magnon signal on RH-ACMs with 100 nm steps while keeping a field of -250 mT along z-direction and a fixed BLS detection frequency of \textbf{a} -17 GHz, \textbf{b} -13 GHz,\textbf{c} 17 GHz,\textbf{d} 13 GHz. The white dashed line indicates the location of RH-ACMs.
}
\label{S-FigModes}
\end{figure}

To further explore the magnon channeling effect as shown in Fig.3 in the main text, we extend the 2D-spatial scanning with BLS detection frequency of -17 GHz,  -13 GHz, 17 GHz, and 13 GHz.  The frequencies were selected based on the spectra obtained by scanning along the ACM as shown in Fig. 2 of the main text. Fig. \ref{S-FigModes}a,c show high intensity at the tube area but low intensity at the helix area, demonstrating its tube mode nature. On the contrary, in Fig. \ref{S-FigModes}b, d, the tube area has significantly lower intensity than the helix region, which demonstrates the magnon channeling effect of the helix.

\clearpage

\section{Field dependence of the toroidal moment for a RH-ACM}
\begin{figure}[h]%
\centering
\includegraphics[width=1\textwidth]{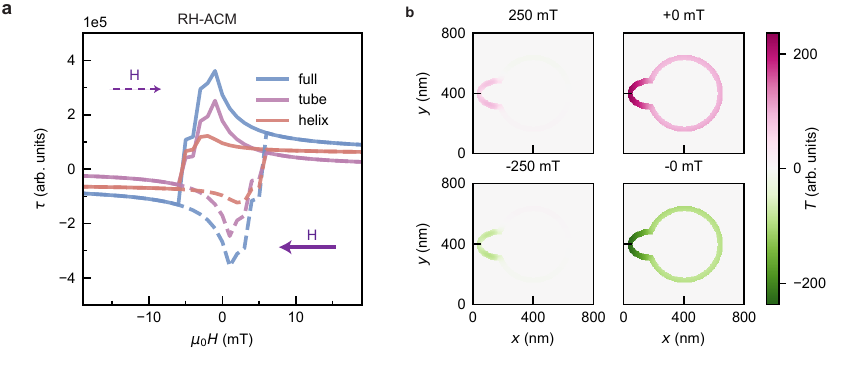}
\caption{\textbf{Field dependence of the toroidal moment of a right handed ACM.}  Toroidal moment $\tau$ of a RH-ACM (blue curve) simulated as a function of axial field $H$ and its sweep direction indicated by arrows. The magenta and orange curves decompose the toroidal moment of the RH-ACM in the contributions from the tube and helix regions, respectively. The field was swept from negative to positive (dashed lines) and back (solid lines). The hysteretic behavior enables a reprogrammable $\boldsymbol{\tau}$ at $H=0$. \textbf{b,} Simulated cross-section of $\tau$ distribution on a RH-ACM when varying the magnetic field from +250 mT, to +0 mT, $-$250 mT and $-$0 mT. The cross-section was taken at the central plane of the ACM. 
}
\label{RH_tormom}
\end{figure}

\clearpage

\section{$\Delta f$ and $\Delta I$ for RH- and LH-ACMs at $\pm$250 mT }
\begin{figure}[h]%
\centering
\includegraphics[width=1\textwidth]{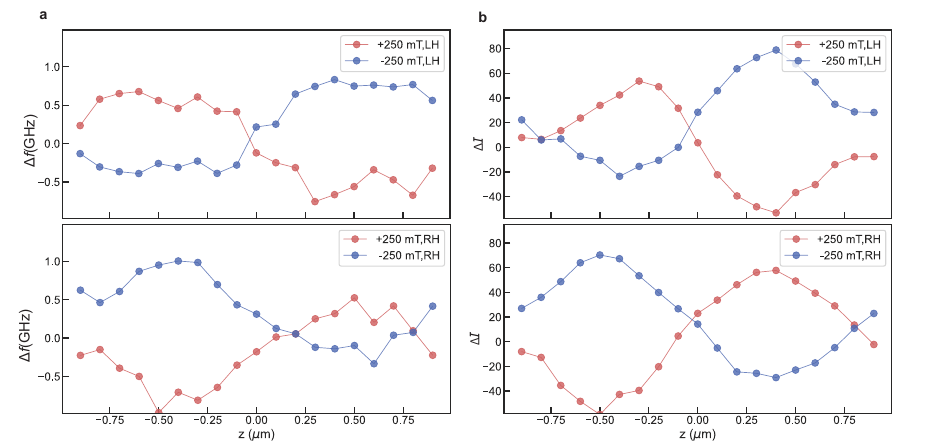}
\caption{\textbf{Frequency difference $\Delta f$  and intensity  difference $\Delta I$.} Quantitative analysis of the nonreciprocity in terms of the frequency difference $\Delta f$ \textbf{a} and intensity difference $\Delta I$ \textbf{b} for the LH-ACM (top) and the RH-ACM (bottom) as a function of position $z$.}
\label{RH_tormom}
\end{figure}

\section{Simulations on a 4000 nm long ACM}
\begin{figure}[h]%
\centering
\includegraphics[width=1\textwidth]{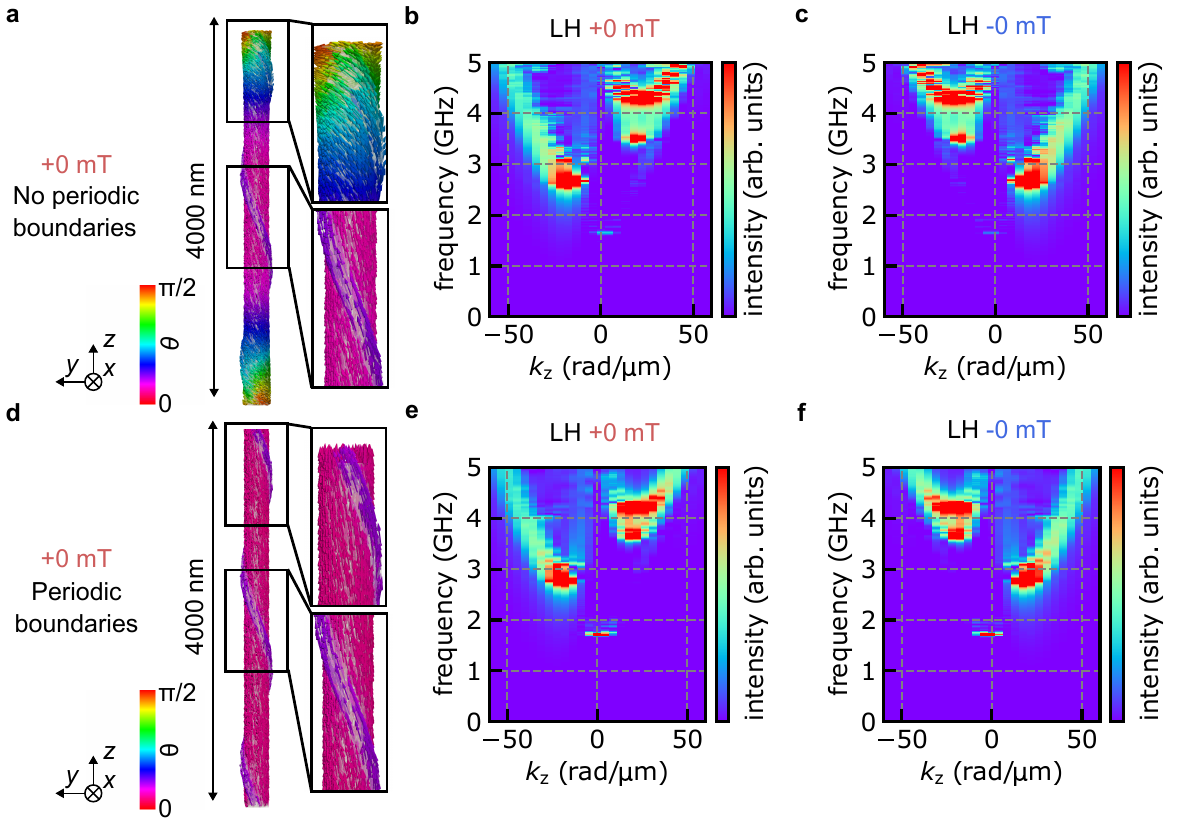}
\caption{\textbf{Dispersion computed for a 4000 nm long ACM} \textbf{a}, Remanent state after saturating along the positive $z$-direction using periodic boundaries. \textbf{b},\textbf{c}, Dispersion for the 4000 nm long ACM with periodic boundaries for remanent states +0 mT and -0 mT, respectively. \textbf{d}, Remanent state after saturating along the positive $z$-direction without periodic boundaries. \textbf{e},\textbf{f} Dispersion for the 4000 nm long ACM without periodic boundaries for remanent states +0 mT and -0 mT, respectively. }\label{fig:supfig4000}
\end{figure}

Simulations were performed on a left-handed 4000 nm long ACM with pitch 2000 nm, tube outer diameter 125 nm, helix outer diameter 200 nm, and thickness 30 nm, according to the same methods as described in the Methods. The damping was quadratically increased to 1 over 800 nm at both ends. The resulting equilibrium magnetization after saturating the structure along the positive $z$-direction is shown in Fig. \ref{fig:supfig4000}a (d) for the case without (with) periodic boundaries. In Fig. \ref{fig:supfig4000}a, the stray field induced curling of the magnetization due to its finite size is clearly seen at the end of the tubes whereas the magnetization in the center of the structure follows the helix. The resulting dispersion at $\pm$0 mT states is shown in Fig.\ref{fig:supfig4000}b,c. Here, to compute the dispersion only the data from the middle of the tube was used (Fig.\ref{fig:supfig4000}a lower panel). The dispersion of the same structure with periodic boundaries and hence with the magnetization in the helical state throughout the tube (Fig. \ref{fig:supfig4000}d,e,f) is qualitatively very similar to that shown in Fig. \ref{fig:supfig4000}b,c, verifying the validity of the results with periodic boundaries shown in Fig. 5. Based on this dispersion, the structure was excited at +0 mT field with a continuous wave excitation with sinusoidal profile of frequency 3 GHz confined to a 20 nm wide strip in the center. The resulting magnon intensity is strongly nonreciprocal as evidenced in Fig.1e of the main text.

\section{Analytical dispersion}\label{theory}
Here we provide the full equations and parameters used for the semi-analytical results of Figure 5. The equations are based on the model provided by Salazar et al. for the magnon dynamics in nanotubes with helical spin textures [1], up to a minus sign for the magnetochiral stiffness field $\mathcal{A}$ [2]. \\

Consider an infinitely long ferromagnetic nanotube aligned along the $z$-axis with saturation magnetization $M_{\rm s}$, exchange stiffness $A$, gryomagnetic ratio $\gamma$ and inner and outer radii $r_{\rm i}$ and $r_{\rm o}$ respectively, mean radius, thickness and cross-sectional area $R=(r_{\rm i} + r_{\rm o})/2$, $T=r_{\rm o}-r_{\rm i}$ and $S=2\pi RT$. The exchange length is $\lambda_{\rm exc}=\sqrt{\frac{A_{\rm exc}}{\frac{1}{2}\mu_{\rm 0}M_{\rm s}^2}}$. The self-demagnetization fields for thin-shells $T\approx\lambda_{\rm exc}$ are [1]:

\begin{eqnarray}
\mathcal{J}(m, k_z)&=&\frac{\pi}{S} \int_0^\infty dq \frac{q^3}{2\qty(q^2 + k_z^2 )}\qty(\Gamma_m\qty[q])^2 \\
\mathcal{K}(m, k_z)&=&\frac{\pi}{S}\int_0^\infty dq \frac{q^2k_z}{q^2+k_z^2}\Gamma_m\qty[q]\Lambda_m\qty[q]\\
\mathcal{L}(m, k_z)&=&\frac{\pi}{S}\int_0^\infty dq\frac{2qk_z^2}{q^2+k_z^2}\qty(\Lambda_m\qty[q])^2\\
\mathcal{M}(m, k_z)&=&m\frac{\pi}{S}\int_0^\infty dq \frac{2qk_z}{q^2+k_z^2}\Lambda_m\qty[q]\mathcal{I}_m\qty[q]\\
\mathcal{N}(m, k_z)&=&m\frac{\pi}{S}\int_0^\infty dq \frac{q^2}{q^2+k_z^2}\Gamma_m\qty[q]\mathcal{I}_m\qty[q]\\
\mathcal{O}(m, k_z)&=&m^2\frac{\pi}{S}\int_0^\infty dq \frac{2q}{q^2+k_z^2}\qty(\mathcal{I}_m\qty[q])^2
\end{eqnarray}

\noindent where $\mathcal{I}_m\qty(q)=\int_{r_{i}}^{r_{0}} d\rho J_m\qty(q\rho)$, $\Lambda_m\qty(q)=\int_{r_{i}}^{r_{o}} d\rho~\rho J_m\qty(q\rho)$ and $\Gamma_m\qty(q)=\Lambda_{m-1}\qty(q)-\Lambda_{m+1}\qty(q)$ and $J_m\qty(x)$ is the Bessel function of first kind and order $m$. Here, $k_z$ is the wavevector along the tube axis and $m$ denotes the azimuthal mode number.

In the ultra-thin limit where $T\approx\lambda_{\rm exc}\ll R$, the factors can be written in terms of the modified Bessel functions $I_m$ and $K_m$ [2]: 

\begin{eqnarray}
\mathcal{J}(m, k_z)&\approx&  1+RT\pdv{I_m\qty(|k_z|R)}{R}\pdv{K_m\qty(|k_z|R)}{R}\\
\mathcal{K}(m, k_z)&\approx&\frac{k_zRT}{2}\frac{\partial}{\partial R}I_m\qty(|k_z|R)K_m\qty(|k_z|R)\\
\mathcal{L}(m, k_z)&\approx&RTk_z^2I_m\qty(|k_z|R)K_m\qty(|k_z|R)\\
\mathcal{M}(m, k_z)&\approx&mk_zTI_m\qty(|k|R)K_m\qty(|k_z|R)\\
\mathcal{N}(m, k_z)&\approx&\frac{mT}{2}\frac{\partial}{\partial R}I_m\qty(|k_z|R)K_m\qty(|k_z|R)\\
\mathcal{O}(m, k_z)&\approx&m^2\frac{T}{R}I_m\qty(|k_z|R)K_m\qty(|k_z|R).
\end{eqnarray}

In the helical state, the orientation of the dimensionless equilibrium magnetization $\bf{m}$ can be specified by the angle $\theta$ between the magnetization and the $z$-axis. If $\theta$ is limited to the interval $\qty[0,\pi/2]$ with $\theta=0$ corresponding to the axial state and $\theta=\pi/2$ corresponding to the vortex state, the magnetization can be decomposed as:

\begin{equation}
    \bf{m}=\chi \sin \qty(\theta)\hat{e}_\varphi + p\cos \qty(\theta)\hat{e}_z,
    \label{eq:dispersion_supp}
\end{equation}

\noindent with $\chi ,p\in \qty{-1,+1}$ the helicity and polarization respectively and $\hat{e}_{\varphi}$ and $\hat{e}_z$ the azimuthal and axial unit vectors.

The fields $\mathcal{A}, \mathcal{B}, \mathcal{C}$ that show up in the dispersion for spin waves in the helical state (eq. 6 of the main text):

\begin{equation}
    \omega_{m}\left(k_z\right)=\omega_{M}\left[\mathcal{A}_m\left(k_z\right) + \sqrt{\mathcal{B}_m\left(k_z\right)\mathcal{C}_m\left(k_z\right)} \right]
\end{equation}

\noindent can be expressed, assuming the absence of uniaxial anisotropy and applied field, as 

\begin{eqnarray}
\mathcal{A}_{k_{z}}^m&=&-\chi \mathcal{K}\qty(m, k_z)\sin \qty(\theta) + p\qty(\mathcal{N}\qty(m,k_z)-\frac{2m\lambda_{exc}^2}{b^2})\cos \qty(\theta) \\
\mathcal{B}_{k_{z}}^m&=&\lambda_{exc}^2\qty(k_z^2+\frac{m^2}{b^2})+\frac{\lambda_{exc}^2}{b^2}\cos^2\qty(\theta)+\mathcal{J}\qty(m,k_z)\\
\mathcal{C}_{k_{z}}^m&=& \lambda_{exc}^2\qty(k_z^2+\frac{m^2}{b^2})+\frac{\lambda_{exc}^2}{b^2}\cos \qty(2\theta)+\mathcal{O}\qty(m,k_z)\cos^2\qty(\theta)\\
&&-\chi p\mathcal{M}\qty(m,k_z)\sin \qty(2\theta) + \mathcal{L}\qty(m,k_z)\sin^2\qty(m,k_z) \nonumber
\end{eqnarray}

\noindent where we have introduced $\frac{1}{b^2}=\frac{1}{RT}\ln \qty(\frac{R+T/2}{R_T/2})$.\\

\begin{figure}%
\centering
\includegraphics[width=1\textwidth]{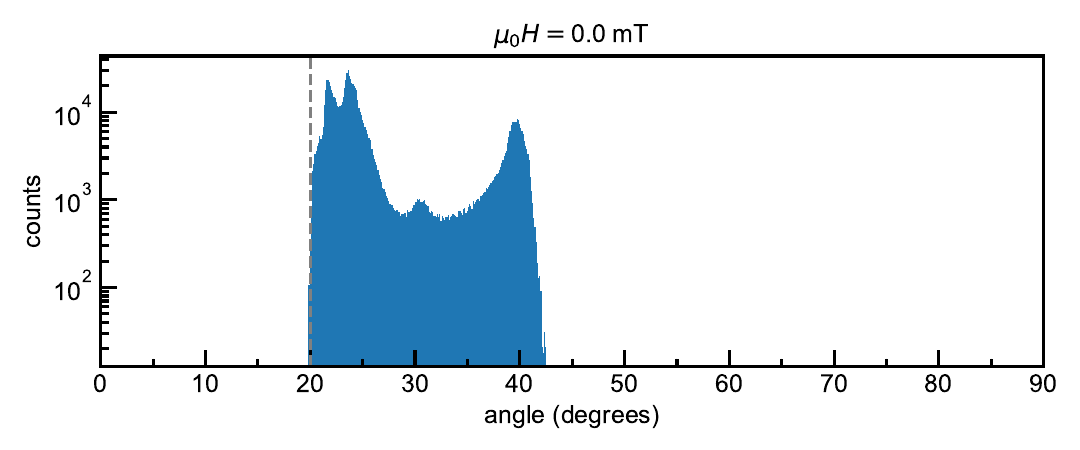}
\caption{\textbf{Numerically computed distribution of the angle $\theta$ at +0 mT of an ACM.} The vertical dashed line indicates an angle of 20$^{\circ}$ which was used as input for the analytical dispersion shown in Fig. 5a of the main text.}\label{fig:supfig1}
\end{figure}

Equation \ref{eq:dispersion_supp} was solved numerically using identical magnetic parameters as were used for the micromagnetic simulations discussed in Fig. 5 of the main text. For comparison with the numerically computed dispersion, the distribution of the magnetization angle $\theta$ of the respective tubes of remanent state was computed in the simulation. In Fig. \ref{fig:supfig1}, we show the distribution for the RH-ACM at +0 mT of Fig. 5a of the main text. Two main peaks can be identified: a peak near 40$^{\circ}$ due to the magnetization of the helix, and a peak at lower angles due to the magnetization in the tubular segment. For the comparison with the numerical simulations, the angle of the tubular segment is most important. Based on the distribution of angles and the quantitative agreement between theory and simulations, we used $\theta=20^{\circ}$.

%For the given geometry, the helix makes an angle $\arctan \qty(\frac{pitch}{2\pi R})$ with the in-plane axis. The corresponding angle with the $z$-axis can be easily extracted from this, which for the given parameters yields $\theta=36.4$
\newpage
\section*{Reference}

[1] M. M. Salazar-Cardona, L. Körber, H. Schultheiss, K. Lenz, A. Thomas, K. Nielsch, A. Kákay, and J. A. Otálora, Nonreciprocity of spin waves in magnetic nanotubes with helical equilibrium magnetization. Applied Physics Letters 118(26), 262411 (2021) https://doi.org/10.1063/5.0048692 . 
\begin{flushleft}
[2] Körber, L.: Spin waves in curved magnetic shells. PhD thesis, Technische Universit at Dresden (2023). https://doi.org/10.25368/2023.131  .
\end{flushleft}

%\bibliography{ACM.bib}
\clearpage